\documentclass{proceedings}
\usepackage{amsmath}

\begin{document}

\title{Thermocapillary Fluid  and Adiabatic Waves\\ Near  The Critical Point}

\author{Henri    Gouin}

\address{
Universit{\'e} d'Aix-Marseille\ \& CNRS\ UMR 6181\\ Av. Escadrille Normandie-Niemen 13397 Marseille Cedex
20, France\ \  (\,\footnote{ Email: henri.gouin@univ-cezanne.fr}\, \footnote{\textbf{Proceedings of  12$^{th}$WASCOM:  \emph{Waves an{d Stability in Continuous Media}, p.p. 253-268, World Scientific Publishing (2004).}}\,}\,)}

\maketitle

\abstracts{Isothermal interfacial zones are investigated starting
from a local energy which can be considered as the sum of two
terms:  one corresponding to a medium with a uniform composition
equal to the local one and a second one associated with the
non-uniformity of the fluid. The additional term can be
approximated by a gradient expansion, typically truncated to the
second order. A representation of the energy near the critical
point therefore allows the study of interfaces of non-molecular
size. Capillary layer and bulk phases are not considered
independently. Obviously, this model is simpler than models
associated with the renormalization-group theory. Nevertheless, it
has the advantage of extending easily well-known results for
equilibrium cases to the dynamics of interfaces. The equation of
state of a one-component system may be expressed as a relation
among the energy, entropy and matter density, $\alpha, s$ and $
\rho$ in the form $\alpha = \alpha(\rho,s)$.  Now, let $\alpha
(\rho,s)$ be the analytic $\alpha$ as it might be given by a
mean-field theory. In the simplest case, in an {\it extended van
der Waals theory\cite{vdW}}, the volume internal energy
$\varepsilon$ is proposed with a gradient expansion depending not
only on $ {\rm grad}\ \rho\ $ but also on ${ \rm grad}\ s $ (the
associated fluid is called thermocapillary fluid\cite{casal}):
\begin{equation}
\varepsilon = f(\rho, s, {\rm grad}\ \rho,   {\rm grad}\ s )
\label{A}
\end{equation}
 With  an energy in the   form (\ref{A}), we
 obtain the
equations of motion of conservative movements for nonhomogeneous
fluid near its critical point. For such a medium, it is not possible
to obtain shock waves. The idea of studying interface motions as
localized travelling waves in a multi-gradient theory is not new
and can be traced throughout many problems of condensed matter and
phase-transition physics\cite{gouin2}. Here, adiabatic waves are
considered and a new kind of waves appears. The waves are
associated with the spatial second derivatives of entropy and
matter density. In Cahn and Hilliard's model\cite{cahn}, the
direction of solitary waves was along the gradient of density. For
this new kind of adiabatic waves, the direction of propagation is
normal to the gradient of densities. In the case of a thick
interface, the waves are tangential to the interface and the wave
celerity is expressed depending on thermodynamic conditions at the
critical point.}
\noindent \emph{Keywords\,:} Exceptional waves; isentropic motions; fluid interfaces; extended van der Waals theory.

\noindent \emph{PACS, MSC Numbers\,:\,64.60.Ht; 74Jxx; 47.35.Fg; 64.70.Fx; 47.35.Pq; 47.55.N-; 46.40.Cd}

\section{Equations of motion of thermocapillary fluids}

\subsection{Conservative  motions }
For conservative motions of perfect fluids -that is to say without
 heat fluxes or viscosity- the specific entropy of each particle remains
 constant along trajectories ($\dot s = 0$, where $\ {\bf \dot {}} \ $ denotes
  the material derivative).
 A convenient method allowing to
 obtain the equations of motion comes from the Hamilton principle.
 The notations are in Serrin\cite{serrin}:
in a fixed coordinate system, the components of a vector
(covector)    $\mathbf{a}$  are  denoted by $a^i $, ($a_i $),
where $i = 1, 2, 3 $. In order to describe the fluid motion
analytically, we refer to the coordinates $\mathbf{x} \equiv\
(x^1,x^2,x^3)$ as the particle's position (Eulerian variables).
The corresponding reference position is denoted by $\mathbf{X}
\equiv\ (X^1,X^2,X^3)$ (Lagrangian variables). The motion of a
fluid is classically represented by the transformation
$\displaystyle \mathbf{x} = {\varphi} (t,\mathbf{X}) \ \mathrm{or}
\ x^i = \varphi^i(t,\mathbf{X}) $, where $t$ is the time. It is
assumed that ${ \varphi } $ possesses an inverse $\mathbf{X} = {
\phi }(t, \mathbf{x}) $ and continuous derivatives up to the
second order except at certain surfaces.\\ We denote by
 $ {\mathbf z} =$
 $ {\left(\begin{array}{c} t \\{\mathbf x} \end{array} \right)}$
the time-space variables; $\displaystyle {\mathbf u
 } \equiv {\partial \varphi\over
\partial t}\ (t,\mathbf{X})$ and ${\mathbf{V}} \equiv
\left(\begin{array}{c} 1 \\ {\mathbf u}
\end{array} \right)$
are respectively the velocity and the time-space  velocity. Let us
consider a mobile surface $\Sigma_t$ defined in the physical space
$D_t$ occupied by the fluid. Let us denote by $g$ the celerity of
$\Sigma_t, \ {\mathbf n}$ its
normal vector,  $\mathbf{N} = \left(\begin{array}{c} -g \\
{\mathbf n}
\end{array} \right)$ and
$v = {\mathbf{N}}^* {\mathbf{V}} \equiv
{\mathbf{n}}^*{\mathbf{u}}- g$   the fluid velocity with respect
to $\Sigma_t$, where $^*$ is the transposition on $D_t$.\\ The
Lagrangian
 of a thermocapillary fluid is written
$$
 L =   {1\over 2}\ \rho {\mathbf u}^* {\mathbf u}  - \varepsilon - \rho \Omega,
$$
where  $\Omega$ is the extraneous force potential defined  as a
function of $\mathbf z$ and $\varepsilon$ is the internal energy
per unit volume. Between times $t_1$ and $t_2$, Hamilton's action
is
 $$
 a = \int_{t_1}^{t_2}\int_{D_t} L\ dv dt
 $$
 A variation of the particle motion comes from a family of virtual motions
 $ {\mathbf X} =
\psi(t, {\mathbf x},r)$ where   $r$ denotes a small
 parameter close to $0$.
 The real motion  is associated with $r  = 0\  \big(\,
 \psi(t, {\mathbf x},0) = \phi(t, {\mathbf x})\, \big)$
  and a virtual
 displacement is expressed in the form\cite{gouin0}
 \begin{equation}
 \delta {\mathbf X} =
 \left({\partial  \psi  \over \partial r}\right)_{r
 =0}  \label{B}
 \end{equation}
Expression (\ref{B})
 corresponds to the dual of the variation given by  Serrin\cite{serrin}  in
 page 145.
 The density and the specific entropy verify respectively
 \begin{equation}
 \rho\ {\rm det}\, F = \rho_0(X), \label{C1}
  \end{equation}
   \begin{equation}
 s = s_0(X), \label{C2}
  \end{equation}
where $\rho_0$ and  (for conservative motions) $s_0$ are defined
into a reference space $D_0$ and at time $t$ fixed, $F$ is the
Jacobian of $\varphi$. If $\varepsilon$ given by relation
(\ref{A}) is differentiable,
$$\delta \varepsilon = \varepsilon_{,\rho}\ \delta \rho +
\varepsilon_{,s}\ \delta s + \Phi^i\ \delta \rho_{,i} + \Psi^i\
\delta  s_{,i}\,,
$$
and consequently, the theory introduces two new vectors
${\mbox{{\boldmath $\Phi$}}} $ and ${\mbox{{\boldmath $\Psi$}}} $
such that :
$$\Phi^i  = \varepsilon_{,\rho_{,i}}\ \ \ {\rm and} \ \ \ \Psi^i  =
\varepsilon_{,s_{,i}}
$$
(In the case of compressible fluids, scalars $\varepsilon_{,\rho}$
and $(1/\rho)\, \varepsilon_{,s}$ are the specific enthalpy and
the Kelvin temperature).\\ Due to the fact the fluid is isotropic,
${\rm grad}\ \rho$ and ${\rm grad}\ s$ are taken into account by
their scalar products only\cite{serrinbis}. Let us denote
$$
\beta = ({\rm grad}\ \rho)^2, \ \chi = {\rm grad}\ \rho\,.\, {\rm
grad}\ s, \  \gamma = ({\rm grad}\ s)^2
$$
and in variables $\rho, s, \beta, \chi, \gamma,$
$$
\varepsilon = g(\rho, s, \beta, \chi, \gamma)
$$
Consequently,
 \begin{equation}
 {\mbox{{\boldmath
$\Phi$}}} = C\ {\rm grad}\ \rho + D\ {\rm grad}\ s, \hskip 0.5cm
 {\mbox{{\boldmath
$\Psi$}}} = D\ {\rm grad}\ \rho + E\ {\rm grad}\  s, \label{D}
  \end{equation}
  with
  $$ C = 2 \ \varepsilon_{,\beta},\ D = \varepsilon_{,\chi}, \ E =
  2\ \varepsilon_{,\gamma}
  $$
Variations of Hamilton's action are deduced from classical method
of variational calculus\cite{gouin0},
\eject
$$
\delta a = \int_{t_1}^{t_2}\int_{D_t}\Big({1\over 2}\ {\mathbf
u}^* {\mathbf u}\ \delta\rho + \rho  {\mathbf u}^*  \delta{\mathbf
u}  - \varepsilon_{,\rho}\  \delta \rho - \varepsilon_{,s}\ \delta
s\
$$
$$
\hskip 2 cm  \ -\, {\mbox{{\boldmath $\Phi$}}}^*\ \delta\ {\rm
grad}\ \rho - {\mbox{{\boldmath $\Psi$}}}^*\ \delta\ {\rm grad}\ s
- \Omega \ \delta \rho \Big)\ dv  dt
$$
 Due to the
definition of  virtual displacement in (\ref{B}),
$$
\delta\ {\rm grad}\ \rho = {\rm grad}\ \delta  \rho,\ \   \delta\
{\rm grad}\ s = {\rm grad}\ \delta  s \ \ {\rm and}\ \ \delta
{\mathbf u} = - F\ \dot{\overset \frown{\delta {\mathbf X}}}
$$
Due to the fact the virtual displacement  $\delta {\bf X}$ and its
derivatives are assumed null at the boundary of $D_t$, integration
by parts using Stokes' formula yields
$$
\delta a = \int_{t_1}^{t_2}\int_{D_t} \bigg( \Big( {1\over 2}\
{\mathbf u}^* {\mathbf u} - \varepsilon_{,\rho}\ + {\rm div}\
{\mbox{{\boldmath $\Phi$}}} - \Omega\Big)\  \delta \rho$$
$$
 +
\Big({\rm div}\ {\mbox{{\boldmath $\Psi$}}} - \varepsilon_{,s}\
\Big)\ \delta s +  \rho \ \dot{\overset \frown{{\mathbf u}^* F}}\
\delta {\rm X}\bigg)\ dv dt
$$
Taking into account   relations\cite{gouin0}
 \begin{equation}
\delta \rho = \rho\ {\rm div_0}\ \delta {\mathbf X} + {1\over {\rm
det} F}{ \partial \rho_0 \over \partial {\mathbf X}}\ \delta
{\mathbf X}, \label{E}
\end{equation}
where $\rm div_0$ denotes  the divergence operator in $D_0$ and
\begin{equation}
\delta  s = {\partial s_0 \over
\partial {\mathbf X}}\ \delta   {\mathbf X},
\label{F}
\end{equation}
we get finally
$$
\delta a = \int_{t_1}^{t_2}\int_{D_0} \rho_0 \left(\dot{\overset
\frown{{\mathbf u}^* F}} - {\rm grad_0}\, m - \theta\ {\rm
grad_0}\,
  s_0 \right) \delta{\mathbf X} \ dv_0 dt
$$
where $\rm grad_0$ is the gradient in $D_0$, and
$$\theta = {1\over \rho} \left( \varepsilon_{,s} - {\rm div}\ {\mbox{{\boldmath
$\Psi$}}}\right), \hskip 0.5cm  h = \varepsilon_{,\rho}\ - {\rm
div}\ {\mbox{{\boldmath $\Phi$}}}, \hskip 0.5cm  m = {1\over 2 }\
{\mathbf u}^* {\mathbf u} - h - \Omega
$$
Then, Hamilton's principle yields
$$
\dot{\overset \frown{{\mathbf u}^* F}} = {\rm grad_0}\ m + \theta\
{\rm grad_0}  \ s_0
$$
Let us note that
$$\left( {\mbox{{\boldmath $\Gamma$}}}^* +
{\mathbf u}^* {{\partial {\mathbf u} \over \partial {\mathbf
x}}}\right ) F = \dot{\overset \frown{{\mathbf u}^* F}} ,$$
 where ${\mbox{{\boldmath $\Gamma$}}}$ is the acceleration vector; we obtain the
equation of motion in the form
\begin{equation}
{\mbox{{\boldmath $\Gamma$}}} = \theta \ {\rm grad}\, s - {\rm
grad}(h+\Omega), \label{G}
\end{equation}
which is the extension of relation (29.8) in Serrin\cite {serrin}.
It is easy to prove that by algebraic calculus, this equation  is
equivalent to the balance of momentum\cite{casal,casal1}
\begin{equation}
{\partial \over\partial t}\,(\rho {\mathbf u^*}) + {\rm div}
(\rho\, {\mathbf u} {\mathbf u^*}-\,\sigma) + \rho\, {\rm grad}\,
\Omega\ =\, 0 \label{GG}
\end{equation}
with $ \displaystyle \sigma_i^j = - ( P - \rho\, {\rm div}\,
{\mbox{{\boldmath $\Phi$}}})\, \delta_i^j - \Phi^j\,\rho_{,i}-
\Psi^j\,s_{,i} \ $  where  $\ P = \rho\, \varepsilon_{,\rho}
-\varepsilon$ (in the case of  compressible fluids, $P$ denotes
the pressure).\\ Relation $\dot s = 0$ is equivalent to
 the balance of energy\cite{casal,casal1}
$$
{\partial e\over\partial t}\,  + {\rm div} \big((e - \sigma)
{\mathbf u} \big) -{\rm div}\, {\mathbf U} - \rho\, {\partial
\Omega\over\partial t} =\, 0
$$
with $ \displaystyle {\mathbf U} = {\dot \rho}\, {\mbox{{\boldmath
$\Phi$}}} + {\dot s}\, {\mbox{{\boldmath $\Psi$}}} \, $ and $\, e
=  {1\over 2}\ \rho {\mathbf u}^* {\mathbf u}  + \varepsilon +
\rho \Omega .$

\subsection{Properties of conservation for isentropic thermocapillary fluids }
Conclusions obtained in (\cite{casal0,serrin}) are easily extended
to thermocapillary fluids. They are deduced from Eq. (\ref{G}).
Let us recall the main results: $\displaystyle J = \oint_C
{\mathbf u}^* d{\mathbf x}$  denotes the circulation of the
velocity on a closed curve convected by the fluid. Then,
\begin{equation}
{dJ\over dt} = \oint_C \theta\ ds
\end{equation}
 We obtain:

{\it Kelvin's circulation theorem}: the circulation of
the velocity on a closed and isentropic curve is constant.\\
 For any motion of isentropic thermocapillary fluid, it is possible
 to introduce
 scalar potentials with the following   evolutions\cite{casal0} :
\begin{equation}
\dot \kappa = m \equiv {1\over 2}\, {\mathbf u}^* {\mathbf u}  - h
- \, \Omega,\ \ \ \dot \tau = 0,\ \ \ \dot\xi =0,\ \ \
\dot\varsigma = 0,\ \ \ \dot s = 0, \label{W}
\end{equation}
such that the velocity field is written
\begin{equation}
{\mathbf u} = {\rm grad}\, \kappa + \xi\ {\rm grad}\, s + \tau\
{\rm grad}\, \varsigma\label{X}
\end{equation}
From eqs. (\ref{W}) and (\ref{X}), one deduces the classification
given by Casal\cite{casal0} for conservative flow of perfect
fluids.

{\it Oligotropic motions: \ } they are motions for which
iso-entropy surfaces are  surfaces of vorticity. The circulation
of the velocity on an isentropic closed curve convected by the
fluid is null. Eq. (\ref{X}) yields
$$
{\mathbf u} = {\rm grad}\, \kappa + \xi\ {\rm grad}\, s
$$

{\it Homentropic motions: \  } $s$ is constant in the fluid and
Eq. (\ref{X}) yields
$$
{\mathbf u} = {\rm grad}\, \kappa  + \tau\ {\rm grad}\, \varsigma
$$

{\it The Cauchy theorem} is extended without difficulty:
$$ {d\over dt} \left({{\rm rot}\, {\mathbf u}\over \rho}\right) =
{\partial {\mathbf u}\over {\partial\mathbf x}}\,{{\rm rot}\,
{\mathbf u}\over \rho}
$$
If  $H = {1\over 2}\, {\mathbf u}^* {\mathbf u}  + h + \, \Omega$,
we deduce from Eq. (\ref{G}), the extended Crocco-Vazsonyi
equation for steady thermocapillary fluid motions :
$$
{\rm rot }\,{\mathbf u}\, \times \,{\mathbf u} = \theta\, {\rm
grad}\, s - {\rm grad}\, H
$$
Conservation laws expressed with Kelvin's theorems are associated
to the group of permutation of particles of the same entropy. This
group keeps the equations of motion invariant. As in
(\cite{gouinA2}) it is possible to associate an expression of the
Noether theorem to this group. It is natural to conjecture such
results for general fluids endowed with an internal energy which
is a functional of matter and entropy densities.

\subsection{Isothermal motions }
 Now, let us consider the case when only the total
variation of the total entropy in $D_t$ is zero\cite{casal}
(virtual displacements  $\delta{\mathbf{X}}$ conserve  the total
entropy of $D_t$). Then,
$$
\delta \int_{D_t} \rho s \ dv = 0
$$
There exists a constant Lagrange multiplier $T_0$ such that the
new Lagrangian
$$
L = {1\over 2}\ \rho\,{\mathbf u}^* {\mathbf u} - \varepsilon -
\rho\, \Omega + T_0\ \rho\, s
$$
yields
$$\delta a = \int_{t_1}^{t_2}\int_{D_t}\delta ( {1\over 2}\
 \rho\, {\mathbf u}^* {\mathbf u} - \varepsilon - \rho\, \Omega + T_0\
 \rho\,
 s
 )\ dvdt = 0
$$
and consequently, as in section 1.1,
$$
\delta a = \int_{t_1}^{t_2}\int_{D_t} \bigg( \Big( {1\over 2}\
{\mathbf u}^* {\mathbf u} - \varepsilon_{,\rho}\ + T_0\ s + {\rm
div}\ {\mbox{{\boldmath $\Phi$}}} - \Omega\Big)\  \delta \rho\  +
$$
$$
\hskip 1cm + \Big({\rm div}\ {\mbox{{\boldmath $\Psi$}}}-
\varepsilon_{,s}\ + T_0\ \rho  \Big)\ \delta s + \rho \
\dot{\overset \frown{{\mathbf u}^* F}}\ \delta {\rm X}\bigg)\ dv
dt = 0
$$
Here $\delta s$ is any scalar field in $D_t$ and $\delta \rho$ is
given by relation (\ref{E}). Finally, we obtain
\begin{subequations}
\begin{eqnarray}
\theta &=&T_0,\label{H1} \\
{\mbox{{\boldmath $\Gamma$}}} &=&  -\ {\rm grad}(\mu +\Omega),
\label{H2}
\end{eqnarray}
\end{subequations}
where  $\ \mu = \varepsilon_{,\rho} -T_0\, s - {\rm div}\
{\mbox{{\boldmath $\Phi$}}} \  $ is the {\it chemical potential}
of the thermocapillary fluid. Eqs (\ref{H1}), (\ref{H2}) are the
equations of motion of an isothermal capillary fluid. Let us
remark that  we obtain also {\it Kelvin's circulation theorem}:
The circulation of the velocity on a closed and isotherm curve is
constant.

\section{Liquid-vapor interface near its critical point}
The critical point associated with the equilibrium of two bulks of
a fluid corresponds to   the limit of their coexistence. The
interface between the phases disappears when that point is
reached. The thickness of the interface increases as the critical
point is approached and it becomes infinite when the interface
itself disappears. Much of what has been done on the theory of the
near-critical interface has been within the framework of the van
der Waals theory\cite{vdW}, so much the present understanding of
the properties of those interfaces comes from that theory or from
suitable extended version of it\cite{rowlinson}. As its critical
point is approached, the gradients of densities are then small.
The present point of view (the interfacial region may be treated
as matter in bulk, with local energy density that is that of a
uniform fluid of composition equal to the local one, with an
additional term arising from the non-uniformity, expressed by a
gradient expansion truncated in second order) is then most likely
to be successful and even qualitatively accurate. In the
following, we consider the
 case when
\begin{equation}
\varepsilon = \rho\, \alpha(\rho,s) + {1\over 2}\Big ( C\ ({
grad}\ \rho)^2 + 2 D\   {grad}\ \rho\, . \, {grad}\ s   + E\
({grad}\ s)^2 \Big ) \label{I}
\end{equation}
where $\alpha$ denotes the specific internal energy of the fluid
in uniform composition, $C, D, E$ are constants and $C E - D^2
> 0$.\\ If $ D = 0$ and $ E = 0$, we are back to the Cahn and
Hilliard model of capillarity\cite{cahn}. If not, we deduce
$$
 h = h_0 - (C \ \Delta \ \rho + D\ \Delta \ s),
\hskip 0.5cm \theta =  T - {1\over \rho}\ ( D \ \Delta \ \rho + E\
\Delta \ s),
$$
where  $h_0 \equiv \alpha + \rho\,\alpha^\prime_\rho  $ and
$T\equiv \alpha^\prime_s$  are respectively  the specific enthalpy
and
 the Kelvin temperature of the homogeneous fluid of matter density $\rho$
and specific entropy $s$.  Let us note that vectors
${\mbox{{\boldmath $\Phi$}}}$ and ${\mbox{{\boldmath $\Psi$}}}$
are always given by expression (\ref{D}), but here coefficients C,
D and E are constant.\\
\noindent At phase equilibrium, Eq. (\ref{H1}) is verified when
$T_0$ is the temperature in the bulks. If we neglect the body
forces and denote $\mu_0 \equiv \alpha + \rho\,\alpha^\prime_\rho
- s T_0 $, we obtain
\begin{subequations}
\begin{eqnarray}
C \ \Delta \ \rho + D\ \Delta \ s &=& \mu_0 -   \mu_1 ,\\
D \ \Delta \ \rho + E\ \Delta \ s &=& \rho\ (T - T_0),
\end{eqnarray}
\end{subequations}
where $\mu_1 = \mu(\rho_l, s_l) = \mu(\rho_v, s_v)$ is  constant
($\rho_l, s_l, \rho_v, s_v$ are the densities in the liquid and
the vapor bulks). In one-dimensional problems, $\rho = \rho(y)$
and equations of equilibrium are associated with   the system
\begin{subequations}
\begin{eqnarray}
C\ \rho^{\prime \prime}+ D \ s^{\prime \prime} &=&
\varepsilon^\prime_\rho
- s\, T_0 - \mu_1 , \label{J1}\\
D\ \rho^{\prime \prime}+ E \ s^{\prime \prime} &=&
\varepsilon^\prime_s  - \rho\, T_0 ,\label{J2}
\end{eqnarray}
\end{subequations}
where $^{\prime\prime}$ denotes the second derivative with respect
to the space variable $y$.
 Near the critical point of the fluid, we use the
representation of $\rho\, \alpha(\rho,s)$ in relation (\ref{I}) in
the form
\begin{equation}
\rho\, \alpha = {B\over 2 A^2} \left(\Big(A(\rho-\rho_c)^2 +
\eta\Big)^2+ \eta^2 \right) +\ \mu_c\, \rho + T_c \, \eta -p_c
\label{J3}
\end{equation}
given by Rowlinson and Widom in (\cite{rowlinson})  when $\eta =
\rho s$ is the entropy per unit volume in which $s_c = 0$,  A and
B are two positive constants associated with the critical
conditions and $\mu_c,\ T_c,\ p_c$ are respectively the values of
the chemical potential, the temperature and the pressure at the
critical point.\\ It is easy to verify that this expression is
equivalent
 to the chemical potential $\mu_0$ of a compressible fluid (case when $C = D = E = 0$)
 in the form :
 $$
 \mu_0(\rho,T_0) = \mu_c + B\ (\rho-\rho_c)^3 - A\ (T_0
-T_c)(\rho-\rho_c),
 $$
 where $\mu_c$ is the value of $\mu_0$ for the critical conditions.
\subsection{Asymptotic analysis of system ({16}) near the critical point}
Due to relation (\ref{J3}), system (16) yields, \\

$\displaystyle  C\ \rho^{\prime \prime}+ D \ s^{\prime \prime} =
2\, B\, (\rho-\rho_c)^3 +2\,{B\over A}\, \rho\, s\, (\rho-\rho_c)
+ {B\over A} (\rho-\rho_c)^2\, s $

$\hskip 2.4 cm \displaystyle +\, 2\,{B\over A^2}\, \rho\, s^2 +
(T_c-T_0)\, s + \mu_c-\mu_1,\\ $

$ \displaystyle D\ \rho^{\prime \prime}+ E \ s^{\prime \prime} =
{B\over A}\, \rho\, (\rho-\rho_c)^2 + 2\, {B\over A^2}\, \rho^2\,
s + (T_c-T_0)\, \rho \\ $

\noindent To consider the physical scales associated with the
interfacial sizes, we look at the change of variables
$$
Y = \epsilon\, y,\ \  \rho(y)-\rho_c  = \epsilon^{n_1}\, R(Y) ,\ \
s(y) = \epsilon^{n_2}\, S(Y),
$$
where $0 < \epsilon \ll 1$, and $\, n_1, \, n_2$ are two
positive constants. We suppose the coefficients $ C, D $ and $E$
to have finite, non-vanishing limiting values at the critical
point.
Then the main part of system (16) leads to\\

$\displaystyle \epsilon^2\left( C  \epsilon^{n_1} { d^2 R\over
dY^2} +
 D   \epsilon^{n_2} { d^2S\over dY^2}\right) = \epsilon^{3 n_1}\, 2\,
 B\, \,R^3 + \epsilon^{n_1 +n_2}\, 2\,{B\over A}\, \rho_c\,R\, S\, $

$\displaystyle \hskip 2 cm \epsilon^{2n_1 +n_2}\,{B\over A}\,
R^2\, S + \epsilon^{2n_2}\, 2\,{B\over A^2}\, \rho_c\, S^2 +
\epsilon^{n_2}\,(T_c-T_0)\, S + \mu_c-\mu_1,\\   $

$\displaystyle \epsilon^2\left( D  \epsilon^{n_1} { d^2 R\over
dY^2} +
 E   \epsilon^{n_2} { d^2S\over dY^2}\right)= \epsilon^{2n_1}\, {B\over A}\, \rho_c\, R^2 +
\epsilon^{n_2}\, 2\, {B\over A^2}\, \rho_c^2\, S + (T_c-T_0)
\rho_c $

\noindent
 It is easy to verify that the solution is
associated with $n_1 =1,\, n_2 = 2$ and $\mu_c = \mu_1$.  Near the
critical point, in   densities $\rho,\, s$ and variable $y$,
system (16) leads to an approximation  in the form :
\begin{subequations}
\begin{eqnarray}
C\ \rho^{\prime \prime}  = \varepsilon^\prime_\rho
- s\, T_0 - \mu_1 ,\label{J4}\\
\varepsilon^\prime_s  - \rho\, T_0 = 0 \hskip 1.32cm \label{J5}
\end{eqnarray}
\end{subequations}
By other arguments, this asymptotic analysis gives the same
results as in (\cite{rowlinson}), page 254, where Rowlinson and
Widom compared the magnitudes of different terms in a system
similar to system     (16): the magnitude of $D\,  s^{\prime
\prime} $ is negligible compared with the typical magnitude of
$\,C\,\rho^{\prime\prime}$ and the magnitude of $\,D\,\rho^{\prime
\prime}+ E \, s^{\prime \prime}$ is
negligible with respect to  the magnitude of $\, \varepsilon^\prime_s  - \rho\, T_0$.\\
\subsection{Integration of system (16) in the approximation of the critical point}
 The approximation of the system (16) yields
 Eq. (\ref{J5}) which is equivalent to $\theta = T_0$ and consequently,
\begin{equation}
2\ \rho s = {A^2\over B}\ (T_0 -T_c) - A\
(\rho-\rho_c)^2\label{KK}
\end{equation}
Taking Eq. (\ref{J4}) into account, we obtain
\begin{equation}
C\ \rho^{\prime \prime}  =  B\ (\rho-\rho_c)^3 - A\ (T_c
-T_0)(\rho-\rho_c) + \mu_c -\mu_1 \label{K5}
\end{equation}
and in the following $\mu_1 = \mu_c$ such that Eq. (\ref{K5}) is
the classical equation of the density profile of a liquid-vapor
interface\cite{rowlinson}. Integration of Eq. (\ref{K5}) yields
$${1\over 2}\ C \rho^{\prime 2}= {B\over 4}(\rho-\rho_c)^4 - {A\over 2}(T_c
-T_0)(\rho-\rho_c)^2 + f_0 ,
$$
where $f_0$ is a constant which verifies for a planar liquid-vapor
interface
$$ \displaystyle f_0 = {A^2\over 4\, B}(T_c -T_0)^2, $$ and we obtain
\begin{equation}
{1\over 2}\ C \rho^{\prime 2}= \left( {{\sqrt B}\,\over
2}(\rho-\rho_c)^2 - { A\over 2 {\sqrt B}}\, (T_c -T_0)\right)^2
\label{L}
\end{equation}
Eq. (\ref{L}) yields the profile of matter density in the
interfacial layer\cite{rowlinson}
$$
\rho(y) = \rho_c + {1\over 2}\ (\rho_l-\rho_v)\ {\rm tanh}
({y\over 2\zeta})
$$
with
$$
\rho_l = \rho_c +\Big({A\over B}\,(T_c-T_0)\Big)^{1\over 2},\
\rho_v = \rho_c - \Big({A\over B}\,(T_c-T_0)\Big)^{1\over 2},\
\zeta = \bigg({C\over 2A\ (T_c-T_0)}\bigg)^{1\over 2} ,
$$
where   $\zeta$ is the characteristic length of the interfacial
layer. Moreover, the surface tension  of the interfacial layer is
:
$$\sigma = \int_{-\infty}^{+\infty} C \rho^{\prime 2}(y) \ dy
\equiv {\sqrt C\over 3B }\ \bigg( 2 A (T_c-T_0)\bigg)^{3\over 2}
$$
\section{Weak discontinuity in conservative motions}
\subsection{Conditions of a weak discontinuity}
To the equation  of conservative motions   given by Eq. (\ref{G}),
we have to add the equation of balance of mass
\begin{equation}
{\partial \rho\over \partial t}\ +\ {\rm div} (\rho {\mathbf u}) =
0,  \label{K}
\end{equation}
the equation of conservation of the specific entropy $\ \dot s = 0\ $ and relation (\ref{J3}).\\
Weak discontinuities of  isentropic motions correspond to
$\displaystyle \rho,\ s,\ {\partial \rho \over \partial {\mathbf
x}},\ {\partial s \over \partial {\mathbf x}},$ continuous through
the wave surfaces. As in Hadamard\cite{hadamard}, we denote by
$[\,\,\, ]$ the jump of a tensorial quantity through a surface of
discontinuity $\Sigma_t$. Consequently, with Hadamard's tensorial
framework, there exits two Lagrange multipliers $\lambda_1,
\lambda_2$ such that :
\begin{subequations}
\begin{eqnarray}
\bigg[\, {\partial \rho \over \partial {\mathbf z}}\,\bigg] = 0\
\Rightarrow\ \bigg[\, {\partial\over\partial {\mathbf z}}\,
({\partial \rho \over
\partial {\mathbf z}})^{^*}\,\bigg] = \lambda_1 {\mathbf N} {\mathbf N}^*\
\Rightarrow\ \bigg[\, {\partial\over\partial {\mathbf x}}\,
({\partial \rho \over
\partial {\mathbf x}})^{^*}\,\bigg] = \lambda_1  {\mathbf n} {\mathbf
n}^*,\ \ \ \
\label{M1}\\
\bigg[\, {\partial s \over \partial {\mathbf z}}\,\bigg] = 0\
\Rightarrow\ \bigg[\, {\partial\over\partial {\mathbf z}}\,
({\partial s \over
\partial {\mathbf z}})^{^*}\,\bigg] = \lambda_2 {\mathbf N} {\mathbf N}^*\
\Rightarrow\ \bigg[\, {\partial\over\partial {\mathbf z}}\,
({\partial s \over
\partial {\mathbf z}})^{^*}\,\bigg] = \lambda_2 {\mathbf n} {\mathbf n}^*, \ \ \ \ \label{M2}\\
{\lambda_1 = [\, \Delta \rho\,]\ \ {\rm and} \ \ \lambda_2 = [\,
\Delta s\,] },\hskip 2.5cm \ \ \ \ \label{M3}
\end{eqnarray}
\end{subequations}
From $[{\mathbf V}] = 0$, we deduce $\displaystyle \Big[\,
{\partial {\mathbf V} \over
\partial {\mathbf z}}\,\Big] = {\mbox{{\boldmath
$\Xi$}}}\ {\mathbf N}^*$  with ${\mathbf N}^* = (-g,{\mathbf n}^*) $ and $ {\mbox{{\boldmath $\Xi$}}} =  \left(\begin{array}{c} 0 \\
{\mathbf H}
\end{array} \right)$ is a vector Lagrange multiplier, where ${\mathbf H}$ is a 3-vector of $D_t$.
Then,
$$
[\dot {\mathbf u}] = \displaystyle \Big[\, {\partial {\mathbf V}
\over
\partial {\mathbf z}}\,\Big] {\mathbf V} =   v\ {\mathbf H},\ \ \ {\rm
with}\ \ \ v = {\mathbf N}^*{\mathbf V} \equiv {\mathbf n}^*
{\mathbf u} - g
$$
Here, $v$ denotes the velocity of the fluid with respect to the
wave
surface of acceleration  $\Sigma_t$.\\
Equation of mass conservation (\ref{K}) is equivalent to
$$
{\partial \rho \over \partial {\mathbf z}}\, {\mathbf V} +\rho\,
{\rm Tr}\left({\partial {\mathbf V} \over
\partial {\mathbf z}}\right) = 0,
$$
where Tr denotes the trace operator. Then,
$\displaystyle\bigg[{\rm Tr}\left({\partial {\mathbf V} \over
\partial {\mathbf z}}\right)\bigg] = 0$ and  consequently,
\begin{equation}
{\mathbf n}^*\, {\mathbf H} = 0   \label{N}
\end{equation}
Equation of conservation of entropy $\dot s = 0$ implies
$$ {\partial{\dot s} \over \partial {\mathbf z}} = 0
 \ \Leftrightarrow \ {\partial\over
\partial{\mathbf z}} \left({\partial{ s} \over
\partial {\mathbf z}}\, {\mathbf V}\right) = 0
$$
Then,\ $\hskip 1.5 cm \displaystyle \Big [\, \left({\partial\over
\partial{\mathbf z}}  \Big({\partial{ s} \over
\partial {\mathbf z}}\Big)^*\,\right)^* {\mathbf V} +
 \left({\partial {\mathbf V} \over
\partial {\mathbf z}}\right)^* \left({\partial s\over
\partial {\mathbf z}}\right)^*\, \Big ] = 0 . $\\
Due to the fact $\displaystyle {\partial\over
\partial{\mathbf z}}  \Big({\partial{ s} \over
\partial {\mathbf z}}\Big)^*$ is a symmetric tensor,
$$
\lambda_2\, {\mathbf N} {\mathbf N}^*\, {\mathbf V} + {\mathbf
N}{\mbox{{\boldmath $\Xi$}}}^* \Big({\partial{ s} \over
\partial {\mathbf z}}\Big)^* = 0\ \  \Leftrightarrow\ \
{\mathbf N}\, \left (\lambda_2\, v + {\partial{ s} \over
\partial {\mathbf x}}\, {\mathbf H} \right) = 0
$$
or,
\begin{equation}
\lambda_2\, v + {\partial{ s} \over
\partial {\mathbf x}}\, {\mathbf H} = 0  \label{O}
\end{equation}
From Rankine-Hugoniot condition associated to Eq. (\ref{GG}), we
obtain the compatibility condition: $[-g \, \rho\, {\mathbf u}^* +
\rho\, {\mathbf n}^* {\mathbf u}{\mathbf u}^* - {\mathbf
n}^*\sigma] = 0,$ and the continuity of $ \rho, \rho_{,i}, s,
s_{,i}$, yields $[\,{\rm div}\, {\mbox{{\boldmath $\Phi$}}}\,] = 0
$, or
\begin{equation}
 [\, C \Delta\rho + D\Delta s\,] = 0 \ \
 \Leftrightarrow \ \  C\, \lambda_1 + D\,\lambda_2 = 0  \label{Q}
\end{equation}
Consequently, there exists a scalar Lagrange multiplier
$\lambda_3$ such that
\begin{equation}
 \Big[\ {\partial\over\partial {\mathbf x}}\left( C \Delta\rho +
 D\Delta s\right)\,\Big] = \lambda_3\, {\mathbf n}^*
  \label{R}
\end{equation}
Equation  of motion (\ref{G}) yields
$$
[\, \dot {\mathbf u}\,] = [\,\theta\,]\, {\rm grad}\,s -
\lambda_3\, {\mathbf n},
$$
$$
 \ {\rm or} \ \  \ \ \  \ \rho \,v \,  {\mathbf H} = -  \left( D \lambda_1 +
E \lambda_2 \right)\, {\rm grad}\, s\
 -\rho\, \lambda_3\, {\mathbf n}
$$
By projection on the normal $ {\mathbf n}$ to $\Sigma_t$ and
taking relation (\ref{N}) into account, we obtain
$$
\rho\,\lambda_3 + {\mathbf n}^*\, {\rm grad}\, s\, \left( D\,
\lambda_1 + E \,\lambda_2 \right)  = 0
$$
By projection on the tangent plane to $\Sigma_t$ and taking
relation (\ref{N}) into account, we obtain
\begin{equation}
 \rho\, v\, {\mathbf H} =  - \left( D\,
 \lambda_1
+ E\,\lambda_2 \right)\,{\rm grad_{tg}} s ,  \label{S}
\end{equation}
where $ {\rm grad_{tg}} s$ denotes the tangential part of  $ {\rm
grad} s$  in $\Sigma_t$. Elimination of ${\mathbf H}$ in the
relation (\ref{S}) comes from relation (\ref{O}), and we get
$$
D\, ({\rm grad_{tg}} s)^2\, \lambda_1 + \left(E\, ({\rm grad_{tg}}
s)^2 -\rho\, v^2\right)\, \lambda_2 =0
$$
Consequently, we obtain a system of three linear equations with
respect to the variables $\lambda_1, \lambda_2, \lambda_3,$
\[
\left \{
\begin{array}{c}
C \lambda_1 + D\lambda_2 = 0,  \\ \\
  D\,({\mathbf n}^*\, {\rm grad} s)\,
\lambda_1 + E \,({\mathbf n}^*\, {\rm grad} s)\,
 \lambda_2   + \rho\,\lambda_3 = 0,\\ \\
 D\, ({\rm grad_{tg}}
s)^2\, \lambda_1 + \left(E\, ({\rm grad_{tg}} s)^2 -\rho\,
v^2\right)\, \lambda_2 =0
\end{array}
\right.
\]
The compatibility of these three equations yields
\begin{equation}
 \rho\,  v^2  =  {(CE-D^2)\, ({\rm grad_{tg}} s)^2\over C}  \label{T}
\end{equation}
Scalar $v$ is the celerity of the surfaces for isentropic weak
discontinuity of acceleration in an isothermal fluid interface
near its critical point.
\subsection{Celerity of isentropic waves of acceleration}
\begin{figure}[ht]
\centerline{\epsfxsize=2.2 in\epsfbox{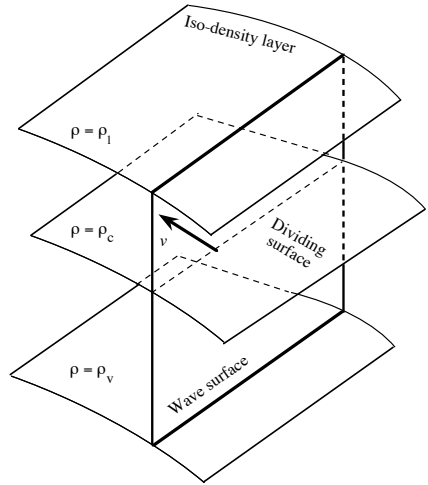}} \caption{The
interfacial layer has a {\it real thickness} of the order of the
characteristic length $\zeta$. The average matter density $\rho_c$
corresponds to a dividing surface, (Rowlinson \& Widom, chapter
3). An isentropic wave surface of acceleration is orthogonal to
iso-density layers and its celerity  of value $v$ is tangential to
the dividing surface.} \label{fig1}
\end{figure}

\noindent The temperature in  liquid and vapor bulks is $T_0$.
Then, relation (\ref{KK}) yields
\[
{\rm grad} \left(\,A (\rho-\rho_c)^2 + 2\, \rho\, s\, \right) =0\
\Leftrightarrow\ A (\rho-\rho_c)\,{\rm grad}\,\rho + {\rm
grad}\,\rho s = 0
\]
The value $\rho=\rho_c$ of the matter density in the interface
corresponds to the maximum value of ${\rm grad}\,\rho$ (see Eq.
(\ref{K5}), when $\mu_c =\mu_1$). The matter density $\rho_c$ is
characteristic of the interfacial matter. For such a value, ${\rm
grad}\,\rho s = 0$ and $s = (A^2/2B \rho_c)\, (T_0-T_c)$.
Consequently, $\displaystyle
 {\rm grad}\,s = {A^2\over  2B \rho_c^2}\, (T_c-T_0)\,  {\rm
 grad}\,\rho,\  {\rm and}
$
$$
  v^2  =  {(CE-D^2)\, A^4\,
  ({\rm grad_{tg}} \rho)^2\, (T_c-T_0)^2\,\over {4\,C\,B^2\,
  \rho_c^5}}
$$
 Due to relation (\ref{L}), we obtain when $\rho=\rho_c,\   $
 $
 \displaystyle C\, ({\rm grad_{tg}}\, \rho)^2 = {A^2\over 2\,B}\,
 (T_c-T_0)^2,
 $  and consequently,
\begin{equation}
  v^2  =  {(CE-D^2)\, A^6\,
   (T_c-T_0)^4\,\over {8\,C^2\,B^3\, \rho_c^5}}  \label{U}
\end{equation}
In the interfacial layer, ${\rm
 grad}\,\rho\,$ is normal to  iso-density surfaces.
 The isentropic waves of acceleration associated with a weak discontinuity
 shear the interfacial layer (see Fig. 1).
The wave celerity, which is proportional to $(T_c-T_0)^2$,
vanishes at the critical point and can be calculated numerically
by means of a state equation.
\section{Results and discussion}
Dynamics of liquid-vapor interfaces is easily studied in mechanics
by means of second gradient theory which is  an extension of the
Landau-Ginzburg models in physics. The theory is associated with
continuous variations of the matter density through the
interfacial layer and was initiated by van der Waals, improved by
Rocard in gas theory\cite{rocard} and Cahn and Hilliard in
physical chemistry\cite{cahn}. Rowlinson and
Widom\cite{rowlinson,widom} pointed out that the model can be
extended by taking into account not only the strong variations of
matter density through the interfacial layer but also the
variations of entropy. Due to the fact the variation of matter
density  leads the variations of entropy (see Eq. (\ref{KK})),
this extension seems at first sight purely formal and yields the
same results that the classical van der Waals model\cite{vdW}
does. Solitary waves, motions… in the normal direction to fluid
interfaces are not involved in an additive dependance of the
entropy gradient. In this paper, we see the dependance of entropy
gradient is necessary for isentropic waves of acceleration along
the interfaces: the fact the internal energy depends not only on
the gradient of matter density but also on the gradient of
entropy, yields a new kind of waves which does not appear in the
simpler models by van der Waals, Cahn and Hilliard, Rocard and
many others\cite{malomed,trusk,slemrod}$^{...}$ : this kind of
waves does not appear when the gradient of entropy is not taken
into account. It is easy to see they are exceptional waves in the
sense of Lax\cite{Ruggeri} and they appear only in, at least,
two-dimension spaces. Recent experiments in space laboratories,
for carbonic dioxide near its critical point have showed the
possibility of such waves.
\section*{Acknowledgments}
This work was performed under the auspices of GDR 2258 CNRS/CNES
by French Spatial Agency. I am grateful to Professor Ruggeri for
his comments and invitation to the ''Wascom 2003'' and to
Professor Gavrilyuk for his criticisms.

\end{document}